\documentstyle[12pt,subeqn,epsf,fullpage]{article}
\def\be{\begin{equation}}
\def\ee{\end{equation}}
\def\bea{\begin{eqnarray}}
\def\eea{\end{eqnarray}}
\def\Ui{\:\raisebox{-1.7ex}{$\stackrel{\textstyle U}{i}$}\:}

\begin{document}
\begin{flushright}
SU-4240-703 \\
TIFR/TH/99-24 
\end{flushright}
\bigskip\bigskip
\begin{center}
{\Large{\bf A QUANTUM ANTI-ZENO PARADOX}} \\[1cm]
{\large A.P. BALACHANDRAN$^{\dagger,\star}$ and S.M. ROY$^{\ddagger,
\star\star}$} \\[1cm] 

$^\dagger$ Department of Physics, Syracuse University,
Syracuse, N.Y. 13244, U.S.A.\\

\bigskip

$^{\ddagger}$ Department of Theoretical Physics, Tata
Institute of Fundamental Research, \\ Homi Bhabha Road, Mumbai 400 005,
India. \\

\vspace{2em}

Abstract 
\end{center}
\bigskip

We establish an exact differential equation for the operator
describing time-dependent measurements continuous in time and obtain a series
solution.  Suppose the
projection operator $E(t) = U(t) E U^\dagger(t)$ is measured
continuously from $t = 0$ to $T$, where $E$ is a projector leaving the
initial state unchanged and $U(t)$ a unitary operator obeying $U(0) =
1$ and some smoothness conditions in $t$.  We prove that the probability of
always finding $E(t) = 1$ from $t = 0$ to $T$ is unity.  If $U(t) \neq
1$, the watched kettle is sure to `boil'.

\vspace{1cm}


\vfill

\noindent $^\star$ E-mail: bal@phy.syr.edu 

\noindent $^{\star\star}$ E-mail: shasanka@theory.tifr.res.in  
\newpage

\noindent {\large 1. \underbar{Introduction}}.  Ordinary quantum
physics specifies probabilities of ideal observations at one instant
of time or of a sequence of such observations at different
instants$^1$.  How should one describe the limit of infinitely
frequent measurements or continuous observation?  One of the earliest
approaches to continuous quantum measurements was already suggested by
Feynman$^2$ in his original work on the path integral.  The Feynman
propagator as modified by measurements is to be calculated by
restricting the paths to cross (or not to cross) certain spacetime
regions (where space can mean configuration space or phase space).  An
approximate way of doing this by incorporating Gaussian cut-offs in
the phase space path integral was developed by Mensky$^3$ who also
showed its equivalence to the phenomenological master equation approach
for open quantum systems using models of system-environment coupling
developed by Joos and Zeh and others$^4$.

On the other hand a completely different approach was initiated by
Misra and Sudarshan$^5$ who asked: what is the rigorous quantum
description of ideal continuous measurement of a projector $E$ (time
independent in the Schr\"odinger representation) over a time interval
$[0,T]$?  Their original motivation$^5$: ``there does not seem to be
any principle, internal to quantum theory, that forbids the duration
of a single measurement or the dead time between successive
measurements from being arbitrarily small'', led them to rigorous
confirmation of a seemingly paradoxical conclusion noted earlier$^6$.
The conclusion ``that an unstable particle which is continuously
observed to see whether it decays will never be found to decay'' or
that a ``watched kettle never boils'' was christened ``Zeno's paradox
in quantum theory'' by Misra and Sudarshan$^5$.  The paradox has been
theoretically scrutinized$^7$ and experimentally probed$^8$.

Here we ask a question far more general than that of Misra and
Sudarshan: what is the operator (the modified Feynman propagator)
corresponding to an ideal continuous measurement of a projection
operator $E_s(t)$ which has an arbitrary (but smooth) dependence on
time in the Schr\"odinger representation?  We obtain a differential
equation for the operator and a series solution which has many   
applications (to be illustrated in a longer paper$^9$).  One of them
leads us to a new watched kettle paradox 
which is apparently quite the opposite of the Zeno paradox, but
mathematically a far reaching generalization of it.  Suppose we
continuously measure from $t = 0$ to $T$ the projector $E_s(t) = U(t)
E U^\dagger(t)$ where $U(t)$ is a unitary operator obeying $U(0) = 1$
and some smoothness conditions, and $E$ a projector obeying $E
\rho(0)E = \rho(0)$, where $\rho(0)$ is the initial density operator.
Then the probability of always finding $E_s(t) = 1$ from $t = 0$ to
$T$ is unity. For the Misra-Sudarshan case $U(t) = 1$ we recover the
usual Zeno paradox that the watched kettle does not boil.  Generically
$U(t) \neq 1$. Hence, for most ways of watching $(U(t) \neq 1)$, the
watched kettle is sure to `boil', an anti-Zeno paradox. If the system
is in an eigenstate of $E$ with eigenvalue unity at $t=0$, it will
change its state with time so as to be in an eigenstate of $E_s(t)$
with eigenvalue unity at all future times.

Our computation of modified Feynman propagators corresponding to
continuous measurements is in the framework of ordinary quantum
mechanics. Exactly the same mathematical expressions for the
propagators, albeit with a different physical meaning would arise in
the `consistent histories' or `sum over histories' quantum mechanics
of closed systems$^{10,11}$, where there is no notion of
measurement. Our computations can therefore be applied also to these
history extended quantum mechanics provided that the probability sum
rules corresponding to consistency conditions or decoherence
conditions are obeyed.  
\bigskip

\noindent {\large 2. \underbar{Formulation of the Problem}}:  For a
quantum system with a self-adjoint Hamiltonian $H$, an initial 
state vector $|\psi(0)\rangle$ evolves to a state vector $|\psi(t)\rangle$,
\be
|\psi(t)\rangle = \exp(-iHt)|\psi(0)\rangle.
\label{one}
\ee
More generally, an initial state with density operator $\rho(0)$ has
the Schr\"odinger time evolution
\be
\rho(t) = \exp(-iHt) \rho(0) \exp(iHt),
\label{two}
\ee
which preserves the normalization condition ${\rm Tr} ~\rho(t) = 1$.
In an ideal instantaneous measurement of a self-adjoint projection
operator $E$, the probability of finding $E = 1$ is ${\rm Tr} (E\rho E)$
and on finding the value $1$ for $E$ the state collapses according to
\be
\rho \rightarrow \rho' = E\rho E/{\rm Tr} (E\rho E).
\label{three}
\ee
If projectors $E_1,E_2,\cdots,E_n$ are measured at times
$t_1,t_2,\cdots,t_n$ respectively, with Schr\"odinger evolution in
between measurements, the probability $p(h)$ for the sequence of
events $h$, 
\be
h: ~E_1 = 1 ~{\rm at} ~ t = t_1; ~E_2 = 1 ~{\rm at}~ t = t_2; \cdots;
~E_n = 1 ~{\rm at}~ t = t_n
\label{four}
\ee
is$^1$
\be
p(h) = ||\psi_h(t')||^2, ~\psi_h(t') = K_h (t') \psi(0), \ t' > t_n.
\label{five}
\ee
Here $K_h (t')$ is the Feynman propagator modified by the events $h$
\be
K_h (t') = \exp(-iHt') A_h(t_n,t_1)
\label{six}
\ee
where,
\be
A_h (t_n,t_1) = E_H(t_n) E_H(t_{n-1}) \cdots E_H(t_1) = T
\prod^n_{i=1} E_H (t_i),
\label{seven}
\ee
with $T$ denoting `time-ordering' and the Heisenberg operators
$E_H(t_i)$ are related to the Schr\"odinger operators by the usual
relation 
\be
E_H (t_i) = \exp(iHt_i) E_s(t_i) \exp(-iHt_i), \ E_s(t_i) \equiv E_i.
\label{eight}
\ee
The state vector of the system at a time $t'$ after the events $h$ is
\[
\psi_h (t') /||\psi_h(t')||.
\]
Correspondingly, if the initial state is a density operator
$\rho(0)$, the probability $p(h)$ for the events $h$ is given by
\be
p(h) = {\rm Tr} ~K_h(t') \rho(0) K^\dagger_h (t') = {\rm Tr} ~A_h
(t_n, t_1) ~\rho(0) A^\dagger_h (t_n, t_1),
\label{nine}
\ee
and the state at $t' > t_n$ is
\[
K_h (t') \rho(0) K^\dagger_h (t')/{\rm Tr} ~(K_h(t') \rho(0)
K^\dagger_h (t')).
\]
\bigskip

\noindent {\large \underbar{Continuous Measurements}}.  Consider
infinitely frequent measurements of the
projection operators $E_s(t_i)$ which are values at times 
$t_i$ of a projection valued function $E_s(t)$. We make the technical
assumption that the corresponding Heisenberg operator $E_H (t)$ is
weakly analytic. We seek to calculate the modified Feynman propagator 
\be
K_h(t') = \exp(-iHt') A_h(t,t_1),
\label{ten}
\ee
where
\be
A_h(t,t_1) = \lim_{n\rightarrow\infty} T \prod^n_{i=1} E_H(t_1 +
(t-t_1) (i-1)/(n-1))
\label{eleven}
\ee
which is the $n\rightarrow\infty$ limit of Eq. (\ref{seven}) with a
specific choice of the $t_i$.  Let us also introduce the projectors
$\bar{E_i} = 1 - E_i$ which are the orthogonal complements of the projectors
$E_i$,
and a sequence of events $\bar h$ complementary to the sequence $h$,
\be
\bar h : \bar{E_1} = 1 ~{\rm at}~ t = t_1; ~\bar{E_2} = 1 ~{\rm at} ~t
= t_2, \cdots, \bar{E_n} = 1 ~{\rm at} ~ t = t_n.
\label{twelve}
\ee
Corresponding to Eqs. (\ref{six}), (\ref{seven}), (\ref{ten}),
(\ref{eleven}), we have Eqs. with $E \rightarrow \bar E$, $h
\rightarrow \bar h$.  
The special interest in $K_{\bar h} (t')$ is that it is closely
related to the propagator 
\be
K_{h'} (t') \equiv \exp(-iHt') - K_{\bar h} (t') = \exp(-iHt') [1 -
A_{\bar h} (t,t_1)] , \ h' \equiv \Ui E_i ,
\label{thirteen}
\ee
which represents the modified Feynman propagator corresponding to the
union of the events $E_i$, i.e. to at least
one of the events $E_s (t_i) = 1$ occurring, with $t_i$ lying between
$t_1$ and $t$.  Our object is to obtain exact operator expressions for
the propagators $K_h$, $K_{\bar h}$ which have been defined above by
formal infinite products. These
results will also provide evaluations of the path
integral formulae for the propagators in history extended quantum
mechanics$^{10,11}$. 
\bigskip

\noindent {\large 3. \underbar{Differential Equation and Series
Solution.}}  We see from Eq. (\ref{ten}) that  $A_h(t,t_1)$ $(A_{\bar
h}(t,t_1))$ represents the modification of the Feynman propagator due to 
the continuous measurement corresponding to the
sequence of events $h(\bar h)$.  Consider first the operators
$A_h(t_i,t_1), ~A_{\bar h}(t_i,t_1)$ before taking the $n \rightarrow
\infty$ limit, and note that 
\be
A_h (t_i,t_1) = E_H (t_i) A_h(t_{i-1},t_1), \ A_{\bar h} (t_i,t_1) =
\bar{E_H} (t_i) A_{\bar h} (t_{i-1},t_1) .
\label{fourteen}
\ee
The relation $\bar E_H^2 = \bar E_H$
implies $A_{\bar h} (t_{i-1}, t_1) = \bar E_H (t_{i-1}) A_{\bar h}
(t_{i-1}, t_1)$. We thus have  
\be
A_{\bar h} (t_i,t_1) - A_{\bar h} (t_{i-1},t_1) =  
(\bar{E_H} (t_i) - \bar{E_H} (t_{i-1})) A_{\bar h} (t_{i-1},t_1) ,
\label{fifteen}
\ee
and a similar relation for $A_h$.  Dividing by
$t_i - t_{i-1} = \delta t$, taking the limit $n \rightarrow \infty$
(i.e., $\delta t \rightarrow 0$) and assuming that $E_H(t)$ is weakly
analytic at $t=0$ we obtain the differential eqns.,
\be
{dA_{\bar h} (t,t_1) \over dt} = {d\bar E_H(t) \over dt} A_{\bar h}
(t_-,t_1), ~ {d A_h(t,t_1) \over dt} = {dE_H(t) \over dt} A_h(t_-,t_1).
\label{sixteen}
\ee
where the arguments $t_-$ on the right-hand sides indicate that in case
of any ambiguity in defining the operator products the
arguments have to be taken as $t - \epsilon$ with
$\epsilon \rightarrow 0$ from positive values and 
\be
{dE_H(t) \over dt} = i[H,E_H(t)] + \exp(iHt) {dE_s(t) \over dt} \exp(-iHt).
\label{seventeen}
\ee
Further $A_{\bar h} (t,t_1),A_h(t,t_1)$ must obey the initial
conditions
\be
A_{\bar h} (t_1,t_1) = \bar{E_H} (t_1), \ A_h(t_1,t_1) = E_H(t_1).
\label{eighteen}
\ee
The measurement differential equations (\ref{sixteen}) are reminiscent
of Schr\"odinger equation for the 
time evolution operator except for the fact that the operators $d\bar
E_H/dt$, $dE_H/dt$ are hermitian whereas in Schr\"odinger theory the
antihermitian operator $H/i$ would occur. Using the
initial conditions we obtain the explicit solutions,
\be
A_h(t,t_1) = T \exp\left(\int^t_{t_1} dt' {dE_H (t') \over dt'}\right)
E_H (t_1),
\label{nineteen}
\ee
and a similar equation with $h \rightarrow \bar h$, $E_h \rightarrow
\bar E_h$, where the time ordered exponentials have the series expansion
\be
T \exp\left(\int^t_{t_1} dt' {dE_H (t') \over dt'}\right) = 1 +
\sum^\infty_{n=1} \int^t_{t_1} dt'_1 \int^{t'_1}_{t_1} dt'_2 \cdots
\int^{t'_{n-1}}_{t_1} dt'_n T \prod^n_{i=1} {dE_H(t'_i) \over dt'_i}.
\label{twenty}
\ee
In general the time-ordered operator products appearing on the
right-hand side are distributions and the
series on the right-hand side must be taken as the definition of the
exponential on the left-hand side; we may not do the integral of
$dE_H(t')/dt'$ on the left-hand side.  Multiplying the expressions for $A_{\bar
h}(t, t_1)$ and $A_h (t, t_1)$ on the left by $\exp (-iHt')$ then
completes the evaluation of the modified Feynman propagators $K_{\bar
h} (t')$ and $K_h (t)$.  
\bigskip

\noindent {\large 4. \underbar{Quantum Anti-Zeno Paradox.}}  We recall
first the usual Zeno paradox. Let the initial state be 
$|\psi_0>$ and let the projection operator $|\psi_0 ><\psi_0|$ be
measured at times $t_1, t_2, \cdots t_n$ with $t_j - t_{j-1} = (t_n -
t_1)/(n-1)$ and $t_n = t$, and let $n \rightarrow \infty$. 
Then, the definition (7) yields,
\bea
A_h (t,t_1) &=& \lim_{n\rightarrow \infty} e^{iHt} |\psi_0> <\psi_0|
\exp (-iH (t-t_1)/(n-1)) |\psi_0>^{n-1} <\psi_0| e^{-iHt_1} \nonumber
\\ [2mm]
&=& \exp (i (H - \bar H) t) |\psi_0> <\psi_0| \exp (-i (H - \bar H)
t_1) ,
\label{twentyone}
\eea
where $\bar H$ denotes $<\psi_0|H|\psi_0>$ and we assume that$^{12}$
$<\psi_0| \exp (-iH\tau)|\psi_0>$ is analytic at $\tau = 0$. Our
differential eqn. also yields exactly this solution for $A_h (t,
t_1)$. Taking $t_1 = 0$, we deduce that the probability $p(h)$ of
finding the system in the initial state at all times upto $t$ is given
by
\be
p(h) = ||K_h (t) |\psi_0> ||^2 = ||\bar e^{i\bar H t} |\psi_0> ||^2 =
1 ,
\label{twentytwo}
\ee
which is the Zeno paradox. (The result can also be generalized to the
case of an initial state described by a density operator, and the
measured projection operator being of arbitrary rank but leaving the
initial state unaltered, see below.) 

\noindent\underbar{Anti-Zeno Paradox}: The above result may suggest
that continuous observation inhibits change of state. Now we prove a
far more general result which shows that a generic continuous
observation actually ensures change of state. Suppose that the initial
state is described by a density operator $\rho (0)$, and we measure
the projection operator
\be
E_s (t') = U (t') E U^\dagger (t') 
\label{twentythree}
\ee
continuously for $t' \epsilon [0, t]$. Here $E$ is an arbitrary
projection operator (which need not even be of finite rank) which
leaves the initial state unaltered,
\be
E \rho (0) E = \rho (0) ,
\label{twentyfour}
\ee
and $U (t')$ is a unitary operator which coincides with the identity
operator at $t' = 0$,
\be
U^\dagger (t') U(t') = U (t') U^\dagger (t') = 1\!\!\!1, U (0) =
1\!\!\!1 .
\label{twentyfive}
\ee
The Heisenberg operator $E_H (t')$ is then
\be
E_H(t') = V(t') E V^\dagger (t'), \ \ V(t') = e^{iHt'} U(t') .
\label{twentysix}
\ee
Clearly $V(t')$ is also a unitary operator. The definition (7) yields,
for $t_1 \geq 0$, 
\be
A_h (t_n, t_1) = V(t_n) (T \prod^{n-1}_{i=1} X (t_i)) V^\dagger
(t_i), \ n \geq 2 
\label{twentyseven}
\ee
where
\be
X(t_i) \equiv E V^\dagger (t_{i+1}) V (t_i) E ,
\label{twentyeight}
\ee
and $A_h (t_1, t_1) = V(t_1) E V^\dagger (t_1)$. Denoting
\be
Y (t_j) = T \prod^{j-1}_{i=1} X(t_i) , \ j \geq 2 , 
\label{twentynine}
\ee
$Y(t_1) = E$ and noting that $E Y(t_{j-1}) = Y(t_{j-1})$, we have

\be
Y(t_j) - Y(t_{j-1}) = E (V^\dagger (t_j) V(t_{j-1}) - 1) E Y(t_{j-1})
.
\label{thirty}
\ee
Taking $t_{j-1} = t', \ t_j = t' + \delta t, \ n \rightarrow \infty$, we have
$\delta t = 0 (1/n)$, and 
\be
E (V^\dagger (t'+\delta t) V(t') - 1) E = \delta t E {dV^\dagger
(t')\over dt'} V(t') E + 0 (\delta t)^2 .
\label{thirtyone}
\ee
To derive that the last term on the right-hand side is $0 (\delta
t)^2$ in the weak sense (i.e., for matrix elements between any two
arbitrary state vectors in the Hilbert space), we make the smoothness
assumption that $E (V^\dagger (t'+\tau) V (t') -1) E$ is analytic in
$\tau$ at $\tau = 0$ in the weak sense. (It may  be seen that this
reduces to analyticity of $<\psi_0| \exp (-iH\tau) |\psi_0>$ in the
usual Zeno case$^{12}$). Hence the $n \rightarrow \infty$ limit yields, 
\be
A_h (t, t_1) = V (t) Y (t) V^\dagger (t_1) ,
\label{thirtytwo}
\ee
where
\be
{dY(t') \over dt'} = E {dV^\dagger (t') \over dt'} V (t') E Y (t') .
\label{thirtythree}
\ee
Solving the differential eqn. we obtain, 
\be
A_h (t, t_1) = V(t) T \exp (\int^t_{t_1} dt' E {dV^\dagger (t') \over
dt'} V (t') E) E V^\dagger (t_1) .
\label{thirtyfour}
\ee
It is satisfying to note that this expression indeed solves our basic
differential equation (16) as can be verified very easily by direct
substitution.

The most crucial point for deriving the anti-Zeno paradox is that the
operator 
\[
T \exp (\int^t_{t_1} dt' E {dV^\dagger (t') \over dt'} V (t') E)
\equiv W (t, t_1) 
\]
is unitary, because $(dV^\dagger (t')/dt')V(t')$ is anti-hermitian as 
a simple consequence of the unitarity of $V (t')$.
Taking $t_1 = 0$, Eq. (\ref{nine}) gives the probability of finding $E_s (t') =
1$ for all $t'$ from $t' = 0$ to $t$ as 
\be
p(h) = {\rm Tr} \left(V(t) W(t,0) E V^\dagger (0) \rho (0) V(0) E
W^\dagger (t,0) V^\dagger (t)\right) = {\rm Tr} \rho (0) = 1 ,
\label{thirtyfive}
\ee
where we have used $V(0) = 1$, $E\rho (0) E = \rho (0)$, the unitarity
of $V(t)$ and the unitarity of $W(t,0)$. This completes the
demonstration of the anti-Zeno paradox: continuous observation of $E_s
(t) = U(t) E U^\dagger (t)$ with $U(t) \neq 1\!\!\!1$ ensures that the
initial state must change with time such that the probability of
finding $E_s (t) = 1$ at all times during the duration of the
measurement is unity.
\bigskip

\noindent \underbar{\large Mathematical remarks.}  The great generality of the 
present results with respect to the ordinary Zeno paradox$^5$ derives
from the fact that the unitary operator $V(t)$ need not obey the
semigroup law$^5$ $V(t) V(s) = V(t + s)$.  Further, the 
following remarks about the set of pairs $(E, \rho)$ [with $\rho$
a density operator] fulfilling $E\rho E = \rho$ can be made. The first
is that as $E$ and $\rho$ are self-adjoint, this condition is
equivalent to either of the requirements $E\rho = \rho$, or $\rho E =
\rho$. They mean just that $\rho$ is zero on the range of $(1\!\!\!1 -
E)$.
The properties of the pairs $(E, \rho)$ in a finite-dimensional
quantum theory are simple. In that case, the density operators, being
a convex set, are connected and contractible while the connected
components of projectors $E$ consist of all the projectors of the same
rank. Thus for fixed rank $n$ of projectors, the allowed pairs $(E,
\rho)$ form a connected space with the structure of a fibre bundle,
with projectors forming the base and a fibre being a convex set. This
bundle is trivial, the fibres being contractible. 
If the quantum Hilbert space ${\cal H}_{n+k}$ is of dimension $n+k$,
its unitary group $U (n+k) = \{U\}$ acts on $(E, \rho)$ by
conjugation: $E \rightarrow U E U^{-1} , \ \rho \rightarrow U \rho U^{-1}$.
This action is an automorphism of the bundle. Since any two projectors
of the same rank are unitarily related, it is also transitive on the
base. 
The nature of the base follows from this remark. The stability group
of $E$ is $U(n) \times U(k)$ where $U(n)$ and $U(k)$ act as identities
on the range of $(1\!\!\!1 - E)$ and $E$ respectively. Thus the base,
as is well-known, is the Grassmannian$^{13}$ $G_{n,k} (C) = U
(n+k)/[U(n) \times U(k)]$. 
When we pass to quantum physics in infinite dimensions, the space of
connected projectors are determined by orbits of infinite-dimensional
unitary groups, and, in addition, a projector can itself be of
infinite rank. In this manner, general applications of our results
will involve infinite-dimensional Grassmannians (on which there are
excellent reviews$^{14}$). 
\bigskip

\noindent \underbar{\large Conclusion.}  It should be stressed that within
standard quantum mechanics and its 
measurement postulates both the usual Zeno paradox and the anti-Zeno
paradox derived here are theorems. The two paradoxes appear
`paradoxical' and `mutually contradictory' only when we forget Bohr's
insistence that quantum results depend not only on the quantum state
but also on the entire disposition of the experimental
apparatus. Indeed the apparatus to measure $E$ and $U(t) E U^\dagger
(t)$ are different.  It would be challenging to see how
these results appear in a quantum theory of closed systems
(including the apparatus) in which there is no notion of
measurements.  It will also be interesting to devise experimental
tests of the anti-Zeno effect along lines used to test the ordinary
Zeno effect$^8$.  

\bigskip

\noindent\underbar{\large{Acknowledgements}} : We would
like to thank Virendra Singh and Rafael Sorkin for discussions. Part
of this work was supported by U.S. DOE under contract
no. DE-FG02-85ER40231. 

\newpage

\noindent {\large \underbar{References}}
\medskip

\begin{enumerate}

\item[{1.}] J. Von Neumann, `\underbar{Mathematical Foundations of
Quantum Mechanics}', Princeton University Press (1955), E.P. Wigner, in
`Foundations of Quantum Mechanics', edited by
B. d'Espagnat (Academic, N.Y. 1971), formulae 14 and 14(a), p.16.

\item[{2.}] R.P. Feynman, Rev. Mod. Phys. \underbar{20}, 367 (1948).

\item[{3.}] M. Mensky, Phys. Lett. \underbar{A196}, 159 (1994).

\item[{4.}] E. Joos and H.D. Zeh, Z. Phys. \underbar{B59}, 223 (1985);
D. Giulini, E. Joos, C. Kiefer, J. Kupsch, I.O. Stamatescu and
H.D. Zeh, `\underbar{Decoherence and the Appearance of a Classical}
\underbar{World}', (Springer-Verlag, Berlin, Heidelberg, N.Y. 1996);
E.B. Davies, `\underbar{Quantum}
\\ \underbar{Theory} \underbar{of Open Systems}', (Academic
Press, N.Y. 1976).

\item[{5.}] B. Misra and E.C.G. Sudarshan,
J. Math. Phys. \underbar{18}, 756 (1977); C.B. Chiu, B. Misra and
E.C.G. Sudarshan, Phys. Rev. \underbar{D16}, 520 (1977);
Phys. Lett. \underbar{B117}, 34 (1982).

\item[{6.}] G.R. Allcock, Ann. Phys. (N.Y.) \underbar{53}, 251 (1969);
W. Yorgrau in `Problems in Philosophy in Science', edited by
I. Lakatos and A. Musgrave (North-Holland, Amsterdam, 1968),
pp.191-92; H. Ekstein and A. Seigert, Ann. Phys. (N.Y.) \underbar{68},
509 (1971).

\item[{7.}] G.C. Ghirardi, C. Omero, T. Weber and A. Rimini, Nuovo
Cim. \underbar{52A}, 421 (1979); H. Nakazato, M. Namiki, S. Pascazio
and H. Rauch, Phys. Lett. \underbar{A199}, 27 (1995).

\item[{8.}] R.J. Cook, Phys. Scr. \underbar{T21}, 49 (1988);
W.H. Itano, D.J. Heinzen, J.J. Bollinger and D.J. Wineland,
Phys. Rev. \underbar{A41}, 2295 (1990); M. Namiki,
S. Pascazio and H. Nakazato, `Decoherence and Quantum Measurements',
(World Scientific, Singapore, 1997); P. Kwiat et al,
Phys. Rev. Lett. \underbar{74}, 4763 (1995); D. Home and
M.A.B. Whitaker, Ann. Phys. \underbar{258}, 237 (1997); M.B. Mensky,
Phys. Lett. \underbar{A257}, 227 (1999). 

\item[{9.}] A.P. Balachandran and S.M. Roy, ``Differential Equation
for Continuous Quantum Measurements and Applications'', TIFR/TH/99-48
in preparation. 

\item[{10.}] R.B. Giffiths, J. Stat. Phys. \underbar{36}, 219 (1984);
M. Gell-Mann and J.B. Hartle, in Proceedings of the XXV$^{\rm th}$
International Conference on High Energy Physics, Singapore, 1990,
Ed. K.K. Phua and Y. Yamaguchi (World Scientific, Singapore, 1991); R.
Omn\'es, J. Stat. Phys. \underbar{53}, 893 (1988); R. Sorkin, in
`Conceptual Problems of Quantum Gravity', Ed. A. Ashtekar and
J. Stachel (Birkhauser, Boston, 1991).

\item[{11.}] J.B. Hartle, Phys. Rev. \underbar{D44}, 3173 (1991);
N. Yamada and S. Takagi, Prog. Theor. Phys. \underbar{86}, 599 (1991).

\item[{12.}] C.B. Chiu, B. Misra and E.C.G. Sudarshan,
Phys. Lett. \underbar{117B}, 34 (1982).   

\item[{13.}] R. Bott and L.T. Tu, ``Differential Forms in Algebraic
Topology'' (Springer-Verlag, 1982); Y. Choquet-Bruhat and
C. Dewitt-Morette, ``Analysis, Manifolds and Physics, Part II:
Applications'' (North-Holland 1989). 

\item[{14.}] A. Pressley and G. Segal, ``Loop Groups'' (Clarendon,
1986); J. Mickelsson, ``Current Algebras and Groups'' (Plenum, 1989). 

\end{enumerate}
\end{document}